\documentclass[lettersize,journal]{IEEEtran}
\usepackage{amsmath,amsfonts}
\usepackage{algorithmic}
\usepackage{algorithm}
\usepackage{array}
\usepackage[caption=false,font=normalsize,labelfont=sf,textfont=sf]{subfig}
\usepackage{textcomp}
\usepackage{stfloats}
\usepackage{url}
\usepackage{verbatim}
\usepackage{graphicx}

\usepackage{cite}

\usepackage{svg}

\usepackage{babel}
\usepackage[T1]{fontenc}
\usepackage[utf8]{inputenc}

\begin{document}

\title{A Mechanistic Model of the \\ Human Menstrual Cycle}

\author{Lena Reitinger, Nina Nikic, Esma Hamzic-Jahic, 
Barbara Arbeithuber, Andreas Springer, Werner Haselmayr, Stefan Angerbauer \\ Institute for Communications Engineering and RF-Systems, Johannes Kepler University Linz, Austria and Department of Gynecology, Obstetrics, and Gynecological Endocrinology, Kepler Universitätsklinikum, Linz, Austria}



\maketitle

\begin{abstract}
The human menstrual cycle is regulated by complex hormonal feedback mechanisms which are essential for reproductive health. Existing mathematical models either rely on simplified phenomenological assumptions or achieve physiological accuracy at the cost of high mathematical complexity and computational costs. This paper presents a novel mechanistic model which balances biological realism with mathematical simplicity. The proposed framework describes the dynamics of the four central cycle-driving hormones, namely follicle stimulating hormone, luteinizing hormone, estradiol, and progesterone, using a system of ordinary differential equations. Mechanistic detail is incorporated through representations of feedback mechanisms, follicular development, luteinization, vascularization of the \textit{corpus luteum}, and aspects of hormone synthesis mechanisms. Despite its comparatively low complexity, the model reproduces characteristic trends of hormone activity and concentration dynamics reported in experimental data. Due to its computational efficiency and modularity, the model is well suited as a basis for extended \textit{in-silico} experiments, including applications in pharmacology and personalized medicine. Our results demonstrate that targeted incorporation of biological mechanisms enables accurate yet tractable modeling of endocrine system dynamics.

\end{abstract}

\begin{IEEEkeywords}
Menstrual cycle, mechanistic model, molecular communication.
\end{IEEEkeywords}

\section{Introduction}

\IEEEPARstart{T}{he} human menstrual cycle is a highly regulated and complex physiological process central to the health and reproductive capacities of about half the human population. Mathematical models have been used for decades to reproduce and investigate the underlying hormone dynamics (e.g., \cite{shack1971mathematical}, \cite{clark2003multiple}) as well as abnormalities and their physiological consequences (e.g., \cite{selgrade2010bifurcation}). Furthermore, \textit{in silico} models of hormone concentration dynamics have been applied to pharmacological use cases, for example, pharmacokinetics (e.g., \cite{roblitz2013mathematical}) or in the context of contraception (e.g., \cite{wright2020mechanistic}, \cite{gavina2023toward}). These applications have resulted in high-dimensional non-linear models, the handling of which is challenging and computationally costly \cite{zhao2024survey}, prohibiting their application in a broader range of research questions. Therefore, the main objective of this presented work is to provide a simple mechanistic model which represents hormone level dynamics throughout the menstrual cycle, while retaining — rather than compromising — mechanistic detail. Advances in the biological understanding of underlying processes nowadays enable an increasingly detailed mechanistic representation, which creates the opportunity to revisit the balance between model complexity and physiological realism. Mechanistic detail helps represent the physiological processes in a more concise manner, i.e., in less complex models, since well-understood interactions can be directly modeled and transformed into mathematical representations. Thus, the goal of this work is to bridge the gap between mechanistic detail and high-dimensional complexity, providing a 10-dimensional ODE-model with 40 parameters which is both physiologically meaningful and practically applicable as a foundation for more comprehensive models.

\begin{figure}
    \centering
    \includegraphics[width=0.5\textwidth]{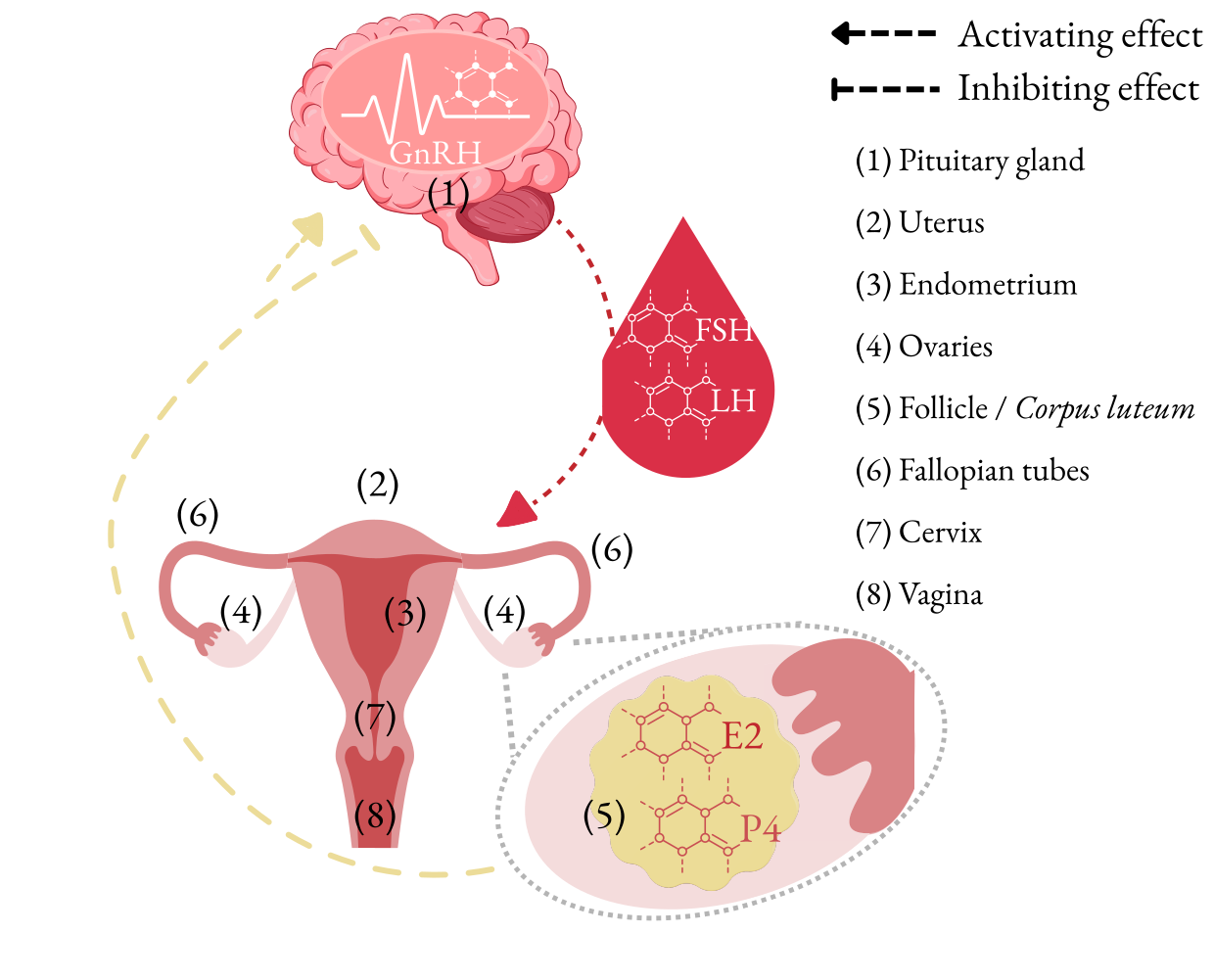}
    \caption{\textbf{The female reproductive system.} The brain and the ovaries are the main sites of hormone synthesis in the menstruating body. The follicle stimulating hormone (FSH) and the luteinizing hormone (LH) are synthesized in the anterior pituitary gland of the brain (under gonadotropin-relasing hormone (GnRH) control) and released to the blood stream. They stimulate processes which cause estradiol (E2) and progesterone (P4) production within the ovaries. FSH, LH, E2 and P4 are the key hormones which drive the menstrual cycle and will be discussed in detail in Sec. \ref{sec:cycle}.} 
    \label{fig:overview}
\end{figure}

\subsection{State-of-the-Art}

The menstrual cycle is a product of monthly recurring patterns of hormone levels. In particular, the follicle stimulating hormone (FSH), luteinizing hormone (LH), estradiol (E2) as well as progesterone (P4) are among the central cycle-driving hormones. Hormone synthesis takes place at different sites in the menstruating body, mostly in hormone-producing glands of the brain and the ovaries in the lower pelvis (Fig. \ref{fig:overview}). Hormone levels fluctuate due to feedback mechanisms which lead to higher or lower hormone synthesis rates (Sec. \ref{sec:cycle}). Hormone interactions are facilitated by the blood stream which allows hormones to travel through the entire body, in particular along the axis between brain and ovaries.

\begin{figure}
    \centering
    \includegraphics[width=0.48\textwidth]{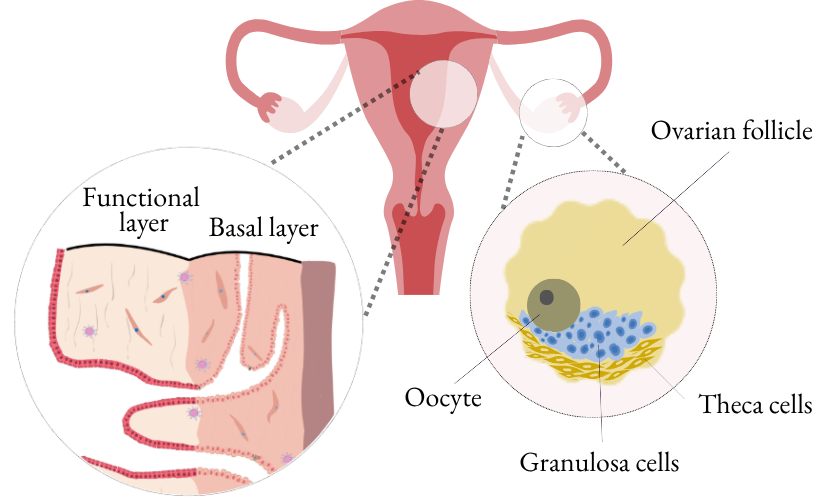}
    \caption{\textbf{Endometrium and ovarian follicle.}} 
    \label{fig:endofollicle}
\end{figure}

Over the past decades, increasingly sophisticated models have been developed, aiming at accurately reproducing hormonal cycles and their regulatory feedback mechanisms. The models typically comprise the interactions of four to nine selected hormones and, depending on the context and purpose of the model, the selection of model components from hormones as well as production sites differ. Early mechanistic models, e.g., \cite{shack1971mathematical}, include the follicle and \textit{corpus luteum} as ovarian sites of hormone synthesis (see Fig. \ref{fig:overview} for their placement within the female reproductive system). Due to the limited knowledge of mechanistic detail at the time, the timing of crucial events of the cycle was modeled with decision functions, hence, these events are initiated once a cascade of prerequisites is met. Since the parameters are estimated based on hormone level data, the simple five-dimensional ordinary differential equation (ODE) model reproduces the dynamics of the four main cycle-driving hormones (FSH, LH, E2 and P4) quite well. With advancing mechanistic understanding and computational resources, more complex and detailed models were proposed. In \cite{clark2003multiple}, which is a combination of \cite{schlosser2000model} and \cite{selgrade1999model}, delay differential equations (DDEs) account for the time delay between a change in the concentration of one hormone and its effect on the synthesis of another. The timing of events in the course of a menstrual cycle is, therefore, becoming a consequence of delayed feedback mechanisms, which is closer to the physiological process. In terms of model components, a fifth hormone is considered in \cite{clark2003multiple}, namely inhibin. The development of the ovarian structures (follicle and \textit{corpus luteum}) is divided into nine stages, the transgression of which is solely regulated by hormone level dynamics. Based on this model, follicular dynamics have been further detailed in \cite{panza2016delay}, where follicular waves, i.e., maturing processes of several follicles in parallel, are assumed to follow each significant increase in FSH concentration, while one of them induces ovulation. Advances in mechanistic detail result in higher complexity of the model, leading to a 13-dimensional non-linear system with 42 parameters in the frame model \cite{clark2003multiple}. Moreover, DDEs induce additional complexity to a model, which affects the computational costs of simulations and challenges the numerical methods to solve these systems. Thus, further progress in mathematical modeling of the menstrual cycle can be generated by substituting DDEs with more mechanistic detail \cite{zhao2024survey}. An approach to this objective is \cite{roblitz2013mathematical}, which includes GnRH dynamics (in the brain, see Fig. \ref{fig:overview} for the physiological context) in addition to the cycle-driving hormones and ovarian hormone synthesis sites as represented in previous models. Because of its application to pharmacological research questions and a generally higher extent of mechanistic realism, the complexity of the resulting model (a 33-dimensional ODE system with 114 parameters) is drastically increased as compared to early models like \cite{shack1971mathematical} despite being DDE-free as well.

In a nutshell, early simple models provide tractable representations of hormonal dynamics but often rely on simplified or phenomenological descriptions of key biological processes, while more recent models improve physiological realism by incorporating additional biological detail and delayed feedback mechanisms, but at the cost of increased mathematical complexity, high dimensionality, and extensive parameterization. This growing complexity can limit interpretability, computational efficiency, and the ability to integrate such models into broader physiological frameworks. Therefore, we aim at improving simple mechanistic models of the menstrual cycle without increasing the complexity of mathematical structures, but instead through a more targeted incorporation of biological processes. More precisely, the main objective is to develop a novel mathematical model of the blood level dynamics of the main cycle-driving hormones FSH, LH, E2 and P4 based on considerably improved mechanistic understanding in order to maintain a comparatively simple and computationally efficient structure. Before proposing the model, however, it is essential to understand the biological background which shall be briefly reviewed below.

\subsection{Contributions}

\begin{itemize}
    \item A novel model of the human menstrual cycle has been developed which is  simpler than existing models in terms of mathematical complexity. The reduction of complexity has been facilitated by a more detailed mechanistic understanding which allows a more concise mathematical representation of the underlying biological processes.
    \item Due to its comparably low computational costs, the model is suitable as a systemic background for mathematical models of further processes in the female reproductive system which are linked to hormone levels (e.g., embryo implantation and endometrial recovery). Furthermore, it can be used as a starting point to address pharmacological and pathology-related research questions.
    \item The proposed model is capable of reproducing trends from published data with the estimated parameters. In a similar fashion, the model is personalizable to the trends of an individual patient.
\end{itemize}

\section{The Menstrual Cycle}
\label{sec:cycle}

In the body of a menstruating person of reproductive age, the endometrium, i.e., the inner lining of the uterus (Fig. \ref{fig:overview} for physiological context and Fig. \ref{fig:endofollicle} for a more detailed depiction), is approximately monthly shed and then rebuilt within the menstrual cycle. This process serves to prepare the body for a potential pregnancy by building a uterine environment where a blastocyst (which develops from a fertilized oocyte, Fig. \ref{fig:endofollicle}) can be implanted, develop into an embryo, and grow further until birth. Unless pregnancy occurs, the uterine environment becomes destructed and the breakdown of the uterine lining results in menstruation. Neatly coordinated interactions of hormones regulate these processes of the menstrual cycle \cite{reece2014campbell}.

\begin{figure}
    \centering
    \includegraphics[width=1\linewidth]{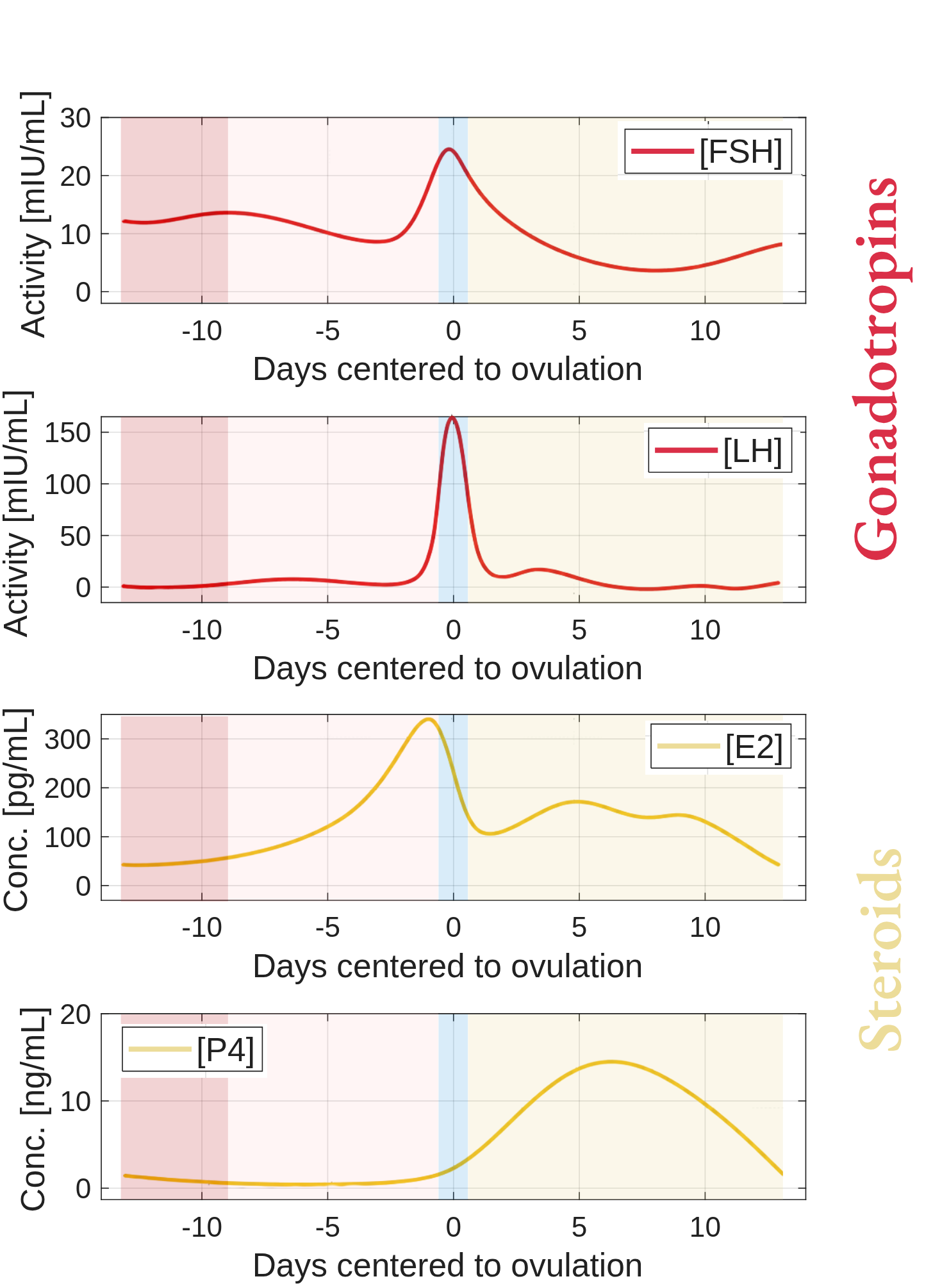}
    \caption{\textbf{Hormone courses over the menstrual cycle.} The figure represents published data on levels of FSH, LH, E2 and P4 in the blood over a cycle from \cite{welt1999female}. Levels of both gonadotropic hormones, FSH and LH, are reported in IU/L. The concentrations of the steroid hormones E2 and P4 are reported in pg/mL and ng/mL, respectively. Cycle phases are indicated with background colour (menstruation in red, follicular phase in rose, ovulation in blue and luteal phase in yellow). We aim at approximating these courses in simulations of the proposed model.}
    \label{fig:scheme}
\end{figure}

\subsection{Phases of the menstrual cycle}

The average human menstrual cycle has a duration of approximately 28 days and is commonly divided in four phases (menstruation, follicular phase, ovulation and luteal phase), according to the central events of the menstrual cycle \cite{reed2015normal}. The phases and corresponding hormone levels are shown in Fig. \ref{fig:scheme}\footnotemark.
\footnotetext{In Fig. \ref{fig:scheme}, the time axis is centered to ovulation due to convention. The start of the first phase of the average 28-days menstrual cycle (i.e., menstruation) therefore corresponds to day -14.}

\vspace{5mm}

\begin{enumerate}

\item \textbf{Menstruation} (approx. 5 days)

The endometrium, i.e., the inner lining of the uterus, consists of two layers: the basal and the functional layer (Fig. \ref{fig:endofollicle}), both of which consist of epithelial, stromal and stem cells. During menstruation, the functional layer is shed, while the basal layer of the endometrium remains. After menstruation, the functional layer starts to regrow from the residual basal layer during the subsequent phases of the menstrual cycle (endometrial regeneration \cite{maruyama2008molecular}). \\

\item \textbf{Follicular phase} (approx. 10 days)

Ovarian follicles contain an immature oocyte which is surrounded by stroma cells (granulosa and theca cells) embedded in a basal membrane \cite{conti2016folliculogenesis}. A schematic view of a follicle is given in Fig. \ref{fig:endofollicle}. Follicles are present within the ovaries during all stages of the menstrual cycle, but during the follicular phase, 5 to 11 follicles enter an advanced stage of maturation. Maturation and growth of these follicles is induced by FSH. Later in the phase, the follicle which responds best to its growth-inducing environment becomes "dominant" and continues its development while the others are degraded \cite{gougeon2010human}. 

In the presence of FSH and the luteinizing hormone (LH), the granulosa and theca cells, respectively, are stimulated and proliferate, leading to an increasing follicle size. Within the follicle, the oocyte is maturing and preparing to leave the follicle in an explosive event, namely ovulation \cite{gougeon2010human}. \\

\item \textbf{Ovulation} (1 day, approx. day 15)

The release of the mature oocyte from the follicle is triggered by a rapid surge in LH concentration. Among other factors, increasing LH levels lead to the weakening and disruption of the follicular wall. The oocyte is released from the ovary to the neighboring fallopian tube where fertilization can take place (Fig. \ref{fig:overview}) \cite{conti2016folliculogenesis}, \cite{richards1998molecular}. \\

\item \textbf{Luteal phase} (approx. 12 days)

After ovulation, granulosa and theca cells of the ruptured follicle undergo a transformative process called luteinization. Under the influence of LH, the morphology as well as the gene expression of these cells change. Firstly, the cell cycle in the remaining follicular cells ceases which halts proliferation and growth. Secondly, altered gene expression leads to a shift in hormone synthesis. Thirdly, vascular endothelial cells proliferate intensely leading to a dense capillary network. This post-ovulatory remainder of the follicle is called \textit{corpus luteum} (CL) and is degraded towards the end of the cycle, unless pregnancy occurs \cite{richards1998molecular}, \cite{devoto2009human}. \\

\end{enumerate}

\begin{figure}[H]
    \centering
    \includegraphics[width=0.5\textwidth]{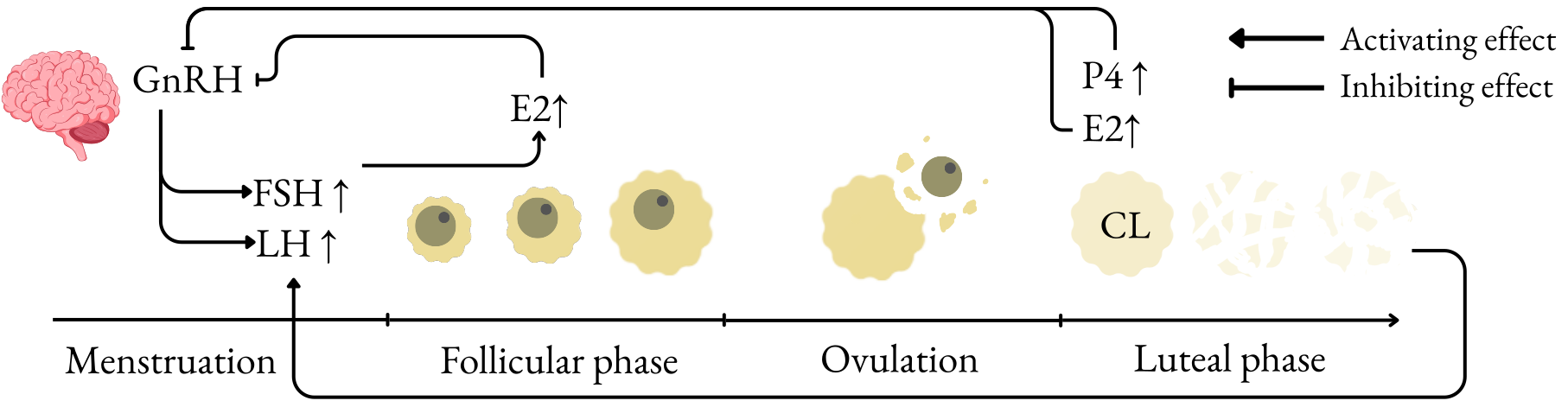}
    \caption{\textbf{Schematic illustration of hormone interactions} during the menstrual cycle.} 
    \label{fig:timeline}
\end{figure}

\subsection{Hormonal interactions}

The cycle-driving hormonal interactions are depicted schematically in Fig. \ref{fig:timeline}. FSH and LH are two central hormones, which are both synthesized in the brain and regulated via the hypothalamus-pituitary gland-ovary axis (HPO axis). More precisely, they are produced within the pituitary gland in response to pulsatile GnRH release patterns at lower or higher frequencies, respectively, from the hypothalamus \cite{marshall1993role}. FSH and LH are termed gonadotropic hormones, since they interact closely with the gonads, i.e., the ovaries in the menstruating body. In the follicular phase, FSH does not only cause the follicles within the ovaries to grow but also to synthesize estrogens, mostly E2. Reversely, E2 exerts an inhibitory effect on FSH and, to a lesser extent, LH synthesis in the brain. As the dominant follicle grows, it produces increasing amounts of E2. As a result of sustained high E2 levels, the feedback mechanism switches, which results in a stimulating effect on FSH and LH production, inducing FSH and LH level peaks \cite{reed2015normal}. Luteinization of the follicular cells leads to a shift in the hormone synthesis machinery within the CL, leading to the production of P4 which, like estrogens, is a steroid hormone. While high amounts of P4 are produced, E2 production does not cease but occurs at a lower level, again resulting in inhibition of LH and FSH synthesis. If no pregnancy occurs in the cycle, P4 levels decrease, the CL is degraded (in a process called luteolysis) and another cycle begins with menstruation and follicular growth. In case of pregnancy, P4 levels are maintained at a high level due to the occurrence of the human chorionic gonadotropin (hCG) \cite{devoto2009human}. \\

\section{Methods}

\begin{figure}
    \centering
    \includegraphics[width=0.45\textwidth]{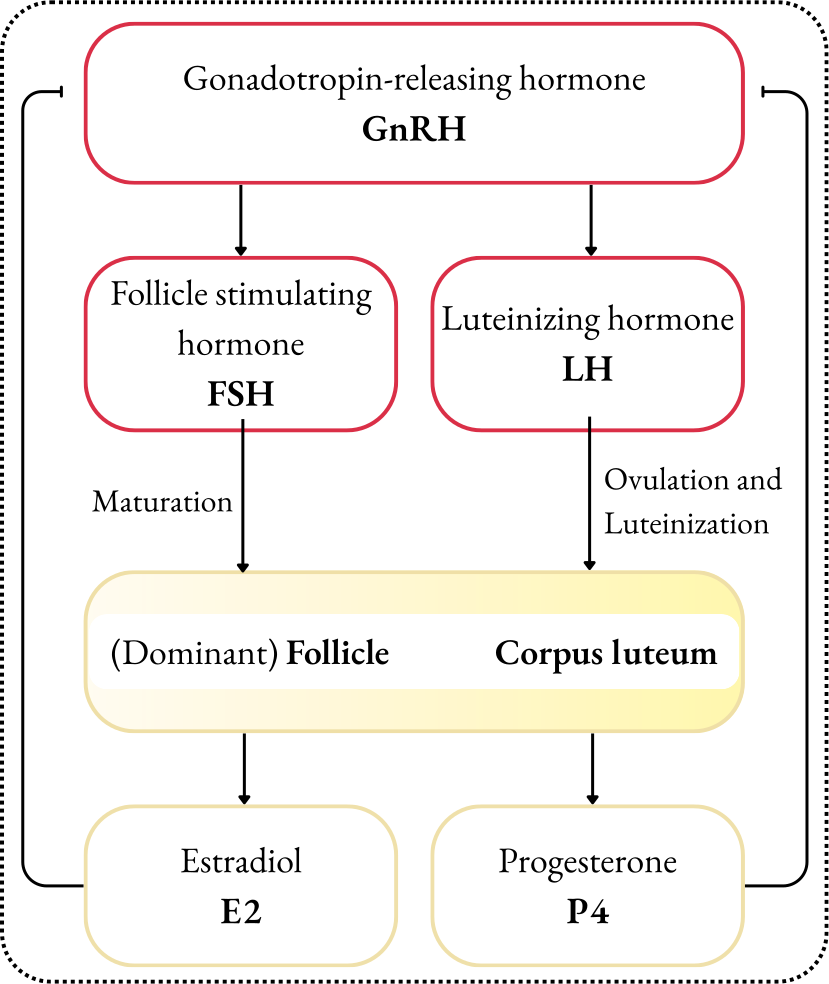}
    \caption{\textbf{Model overview.} The main components of the proposed model are shown: FSH and LH, the synthesis of which is a consequence of GnRH levels in a pulse-frequency encoding fashion. GnRH synthesis is regulated by E2 and P4 feedback. These steroid hormones are produced within the ovarian follicle and its successor, the \textit{corpus luteum} (CL). E2 feedback can be both stimulating and inhibiting GnRH and, hence, FSH and LH synthesis at high and lower levels, respectively.}
    \label{fig:modelscheme}
\end{figure}

The aim of this work is to build a novel mechanistic model which predicts the temporal evolution of concentrations of the four central hormones FSH, LH, E2 and P4 over the menstrual cycle. Importantly, the incorporation of mechanistic detail is intended to keep the mathematical complexity of the model to a minimum, which makes the model a valuable addition to currently existing models. The main model components are depicted in Fig. \ref{fig:modelscheme} and the curves published in \cite{welt1999female} which are intended to be represented by the proposed model are shown in Fig. \ref{fig:scheme}. The corresponding equations are derived in the following subsections.

When developing a mechanistic model, the central objective is to mathematically represent details of underlying biological mechanisms. For that purpose, methods from \cite{ingalls2013mathematical} have been applied to derive systems of ODEs. In particular, Hill functions, which are often used in pharmacological and biological modeling to represent binding events (e.g., of ligands to an enzyme), have been used. Many feedback mechanisms modeled in this work result from the interaction of the effecting hormone (binding molecule) with the corresponding receptors (target molecules) on the surface of, e.g., ovarian cells \cite{bhartiya2021overview} or neuronal cells in the pituitary gland \cite{flanagan2017gonadotropin}. This makes Hill terms of the form $$\frac{\alpha \cdot [\mathrm{A}]^n}{\mathrm{K}^n + [\mathrm{A}]^n} \text{\hspace{2mm}(activating) or \hspace{2mm}} \frac{\alpha \cdot \mathrm{K}^n}{\mathrm{K}^n + [\mathrm{A}]^n} \text{\hspace{2mm} (inhibiting)}$$ particularly useful to represent and adapt the feedback mechanisms in a biologically meaningful manner. Capital letters in square brackets always represent the time-dependent concentration of the respective hormone and $\alpha$ is a hormone synthesis rate. The Hill coefficient $n$ controls the steepness of the sigmoidal curve and $K$ regulates its turning point. The above Hill terms can then be applied to model the synthesis of a hormone X, which is produced at a (constant or time-dependent) rate $\alpha$ dependent on the activating or inhibiting feedback of hormone A. The value of $K$ corresponds to the concentration of hormone A which triggers the turning point in the concentration curve of hormone X. 

\subsection{Gonadotropic hormones (FSH, LH)}
\label{subsec:gonadotropins}

The blood concentrations of the gonadotropins FSH and LH are regulated via the HPO axis. Hence, their secretion depends strongly on the levels of E2, which stimulates or inhibits the synthesis of FSH and LH in the brain \cite{reed2015normal}. The equations \eqref{eq:FSH} and \eqref{eq:LH} represent the resulting FSH and LH dynamics, respectively. 

\begin{equation}
\begin{split}
\frac{\mathrm{d}[\mathrm{FSH}]}{\mathrm{d}t} =  b_{\mathrm{F}} +  \frac{\alpha_{\mathrm{F}_{1}} K_{\mathrm{F}_{1}}^i}{K_{\mathrm{F}_{1}}^i + [\mathrm{E}]^i}+ \frac{\alpha_{\mathrm{F}_{2}} [\mathrm{E}]^i}{K_{\mathrm{F}_{2}}^i + [\mathrm{E}]^i}  \\ + \frac{\alpha_{\mathrm{F}_{3}} K_{\mathrm{F}_{3}}^i}{K_{\mathrm{F}_{3}}^i + [\mathrm{E}]^i + [\mathrm{P}]^i} - k_{\mathrm{F}} [\mathrm{FSH}] 
\end{split}
\label{eq:FSH}
\end{equation}

\begin{equation}
\begin{split}
\frac{\mathrm{d}[\mathrm{LH}]}{\mathrm{d}t} = b_{\mathrm{L}} +  \frac{\alpha_{\mathrm{L}_{1}} K_{\mathrm{L}_{1}}^j}{K_{\mathrm{L}_{1}}^j + [\mathrm{E}]^j} + \frac{\alpha_{\mathrm{L}_{2}} [\mathrm{E}]^j} {K_{\mathrm{L}_{2}}^j + [\mathrm{E}]^j} \\ + \frac{\alpha_{\mathrm{L}_{3}} K_{\mathrm{L}_{3}}^j}{K_{\mathrm{L}_{3}}^j + [\mathrm{E}]^j + [\mathrm{P}]^j} - k_{\mathrm{L}} [\mathrm{LH}]
\end{split}
\label{eq:LH}
\end{equation}

\vspace{5mm}

The initial levels of both hormones are supposed to be greater than zero, since the reported FSH and LH activities are approximately 10 IU/mL at the beginning of a cycle as well as at the end of the previous cycle (Fig. \ref{fig:scheme} \cite{welt1999female}). Therefore, $b_{\mathrm{F}}$ is included as a basal, feedback-independent FSH synthesis rate in \eqref{eq:FSH} and $b_{\mathrm{L}}$ analogously in \eqref{eq:LH} and initial conditions are set for the respective integrators in simulation (Tab. \ref{tab:initials}\footnote{These initial values have been estimated to fit the data reported in \cite{welt1999female} for hormone activities and concentrations and \cite{p1981hormonal} for initial granulosa cell numbers. The initial degree of vascularization has been set to 0.001, since new blood vessels can only form from existing ones which is why this number should be non-zero.}). Furthermore, hormones are supposed to be removed from the blood stream due to degradation at constant rates $k_{\mathrm{F}}$ and $k_{\mathrm{L}}$, which is what the last term of \eqref{eq:FSH} and \eqref{eq:LH} corresponds to.

During menstruation and the follicular phase, increasing E2 levels first lead to an increasing inhibition of gonadotropin secretion, which is represented in the first Hill term of \eqref{eq:FSH} and \eqref{eq:LH}. The parameters $K_{\mathrm{F}_{1}}$, $K_{\mathrm{L}_{1}}$, $i$ and $j$ are the above defined parameters of Hill terms for FSH and LH, and $\alpha_{\mathrm{F}_{1}}$ and $\alpha_{\mathrm{L}_{1}}$ are the synthesis rates of FSH and LH.  The consequences of E2 feedback in this phase are decreasing levels of FSH (after a slight increase before inhibition) and LH in the first 12 days of the cycle (Fig. \ref{fig:scheme}), which is therefore represented in the model, as increasing E2 concentrations ($[\mathrm{E}]$) result in decreasing values of the first Hill terms in both equations.

At high E2 levels, the feedback mechanism is reversed and high E2 levels produce a surge in both LH and FSH. These dynamics are modeled by the second Hill term of \eqref{eq:FSH} and \eqref{eq:LH}, with $K_{\mathrm{F}_{2}}$, $K_{\mathrm{L}_{2}}$, $i$ and $j$ being the parameters of an activating Hill term and $\alpha_{\mathrm{F}_{2}}$ and $\alpha_{\mathrm{L}_{2}}$ the synthesis rates of FSH and LH, respectively. Increasing E2 concentrations ($[\mathrm{E}]$) lead to enhanced synthesis of both FSH and LH, since the value of the term is increasing. Note that $K_{\mathrm{F}_{2}}$ and $K_{\mathrm{L}_{2}}$ are those E2 levels which induce the mid-cycle FSH and LH peaks (Fig. \ref{fig:scheme}).

After the peak, the levels of both gonadotropic hormones decrease due to inhibition as a consequence of then again lower E2 levels. Furthermore, the second central steroid hormone, P4, leads to inhibition of FSH and LH synthesis during the luteal phase. The third Hill term represents these feedback mechanisms with the Hill parameters $K_{\mathrm{F}_{3}}$, $K_{\mathrm{L}_{3}}$, $i$ and $j$, and synthesis rates $\alpha_{\mathrm{F}_{3}}$ of FSH and $\alpha_{\mathrm{L}_{3}}$ of LH. An increasing steroid hormone concentration ($[\mathrm{E}]+[\mathrm{P}]$) yiels a lower value for the described addend.

\subsection{Follicle and \textit{corpus luteum}}

The consideration of the follicle and its luteinized remnant, the CL, is a pecularity of the proposed model. Being major steroid production sites in the follicular and luteal cycle phase, respectively, both are central for the menstrual cycle and its regulation. During the follicular phase, the (dominant) follicle grows due to the fast proliferation of the granulosa and theca cells, while the diameter of the enclosed oocyte remains at around 140 \nolinebreak $\mu$m in these stages of follicular development \cite{gougeon2010human} (see Fig. \ref{fig:endofollicle} for a schematic depiction of the follicle). Folliclular growth is therefore modeled by an initial bunch of stroma cells which pass through the cell cycle phases, namely a first growth phase (gap phase 1, G1), synthesis phase (S), a second growth phase (gap phase 2, G2) and mitosis (M). 

\begin{subequations}
    \begin{alignat}{4}
           \frac{\mathrm{d}G_1}{\mathrm{d}t} &= -\lambda_1 G_1 + 2 \lambda_4 M - d_1 G_1 - r_A G_1,
           \label{G1Growth} \\
            \frac{\mathrm{d}S}{\mathrm{d}t} &= \lambda_1 G_1 - \lambda_2 S - d_2 S - r_A S, \label{S}\\
             \frac{\mathrm{d}G_2}{\mathrm{d}t} &= \lambda_2 S - \lambda_3 G_2 - d_3 G_2 - r_A G_2,  \label{G2Growth}\\
            \frac{\mathrm{d}M}{\mathrm{d}t} &= - \lambda_4 M + \lambda_3 G_2 - d_4  M - r_A M \label{M}
    \end{alignat}
\end{subequations}

\vspace{5mm}

As a cell transgresses these phases, it is doubling in size and cellular components in order to divide into two daughter cells at the end of M phase. The cell cycle dynamics are represented in (\ref{G1Growth})-(\ref{M}), with $\lambda_1$ to $\lambda_4$ being the transgression rates from G1 to S, S to G2, G2 to M and M to G1, respectively, and $d_1$ to $d_4$ the death rates of cells in the respective phase. The cell division which leads to one cell becoming two cells when going through the cycle once is represented in the factor 2 in (\ref{G1Growth}).

As LH levels increase and pass a certain threshold, the stroma cells stop proliferating due to cell cycle arrest \cite{devoto2009human}. In the model, this is represented in a switch-like arrest rate $r_A$ which is zero by default and is set to a positive value once the LH concentration increases over a pre-set threshold ($L_{\mathrm{lim}}$). As a consequence, cells are withdrawn from every cell cycle phase at rate $r_A$ leading to the terms $r_AG_1$, $r_AS$, $r_AG_2$ and $r_AM$ in (\ref{G1Growth})-(\ref{M}). Hence, the cells gradually transition into a new state and equation (\ref{cellcyclearrest}) represents the dynamics of these non-proliferating, hence resting, cells (R), which are degraded at a constant rate $d_R$. Importantly, this transition starts once $[\mathrm{LH}]$ has passed a certain threshold ($L_{\mathrm{lim}}$) when surging. Before this occurs, $r_A$ equals 0 and, therefore, the cells pass through the cell cycle and, thus, proliferate,

\begin{equation}
    \frac{\mathrm{d}R}{\mathrm{d}t} = r_A G_1 + r_A S + r_A G_2 + r_A M - r_L R - d_R R.
    \label{cellcyclearrest}
\end{equation}

With cell cycle arrest and high LH concentrations, another process is initiated: luteinization. Thereby, granulosa and theca cells undergo a reprogramming process which changes their morphology and gene expression \cite{conti2016folliculogenesis}. After ovulation, these luteinized cells constitute the CL, a temporarily existing gland remnant from the follicle. The transformation of non-proliferating cells to luteinized cells is represented in (\ref{luteinization}), with $L$ being the number of luteinized cells, $d_L$ the death rate of luteinized cells and $r_L$ the luteinization rate. When becoming luteinized, cells leave the non-proliferating state at the same rate, which is facilitated by the last term of (\ref{cellcyclearrest}).

\begin{equation}
    \frac{\mathrm{d}L}{\mathrm{d}t} = r_L R - d_L L
    \label{luteinization}
\end{equation}

\subsection{Steroid hormones (E2, P4)}

The follicle- and CL-constituting cells are the main sources of both steroid hormones. E2 and P4 are synthesized from steroid scaffolds in three- and two-step enzymatic reactions, respectively. As to E2, precursor molecules (androgens) are synthesized in theca cells of the follicle and the CL. These steroids are then aromatized in granulosa cells, yielding E2 \cite{hillier1994follicular}. After luteinization, additional enzymes are expressed in theca and granulosa cells which facilitate the synthesis of P4 from its precursor cholesterol. Importantly, cholesterol is mostly retrieved from the blood stream which is why, in addition to luteinization, vascularization (i.e., the development of a blood capillary system) of the CL is a prerequisite of P4 synthesis \cite{devoto2009human}. Since both protein synthesis processes are very short compared to the time scale of the model (days), steady states were assumed for precursor molecules and intermediates \cite{johnson2011original}. The hormone concentrations are derived from the dynamics of their producing cells (E2: granulosa cells, P4: lutein cells), in the case of P4 additionally depending on the degree of vascularization.

\begin{equation}
    \frac{\mathrm{d}[\mathrm{E}]}{\mathrm{d}t} = b_{\mathrm{E}} + \frac{\alpha_{\mathrm{E}_{1}} N_G^i}{K_{{\mathrm{E}_{1}}}^i +  N_G^i} + \frac{\alpha_{\mathrm{E}_{2}} L^i}{K_{{\mathrm{E}_{2}}}^i + L^i} - k_{\mathrm{E}} [\mathrm{E}]
    \label{estradiolsynth}
\end{equation}

With regards to E2, reported initial concentrations are not zero but approximately 50 pg/mL \cite{welt1999female} (Fig. \ref{fig:scheme}), which is why a basal synthesis rate $b_{\mathrm{E}}$ can be found in the first term of \eqref{estradiolsynth} and initial conditions are set according to Tab. \ref{tab:initials}. Furthermore, E2 is degraded at a constant rate $ k_{\mathrm{E}}$, yielding the last term of \eqref{estradiolsynth}. Both Hill terms are activating and include Hill parameters $K_{{\mathrm{E}_{1}}}$, $K_{{\mathrm{E}_{2}}}$ and $i$ as well as the number of granulosa and lutein cells, $N_G$ and $L$ respectively. The former is calculated as the sum of all states of \eqref{G1Growth}-\eqref{M}, i.e., $N_G = G_1 + S + G_2 + M$, and the latter from \eqref{luteinization}. 

In addition to luteinization, vascularization of the CL, i.e., the growth of a blood vessel system within the CL \cite{reynolds2000angiogenesis}, is considered in P4 synthesis,

\begin{equation}
    \frac{\mathrm{d}V}{\mathrm{d}t} = \frac{r_V}{K} (K - V) \; V - d_V V.
    \label{vasc}
\end{equation}

The precursor molecules needed for P4 synthesis are cholesterol molecules circulating in the blood stream, and them reaching the site of P4 production within luteinized cells is crucial \cite{devoto2009human}. The degree of vascularization is, thus, represented in (\ref{vasc}) with $r_V$ determining the vascularization speed, $K$ being the percentage of not yet vascularized cells and $V$ the percentage of the vascularized ones. Importantly, the proportion of vascularized cells is assumed to grow logistically (like the proliferation of vascular cells in \cite{friedman2005analysis}), since new blood vessels are built from proliferating vascular cells and the degree of vascularization is limited due to the constant volume of the CL. Therefore, the rate of vascularization increases first with a growing number of vascular cells which proliferate, hence, create new vascular cells. When a certain degree of vascularization is reached, the rate decreases again, since the CL is packed to capacity with a vascular system. Moreover, vascular cells are dying at a constant rate $d_V$.

P4 is assumed to be synthesized with a constant rate $\gamma_P$ and its levels are modeled following (\ref{Psynt}), since they depend on both the vascularization degree of the CL ($V$) and the number of P4 producing cells ($L$),

\begin{equation}
    [\mathrm{P}] = \gamma_P \; V \; L.
    \label{Psynt}
\end{equation}

\section{Results and Discussion}

\begin{figure}[H]
  \centering
  \includegraphics[width=0.5\textwidth]{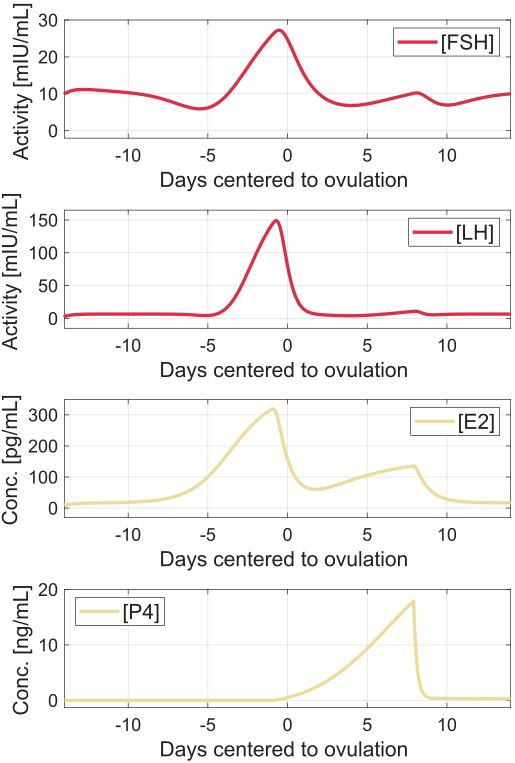}
  \caption{\textbf{Hormone activities and concentrations in an average 28-days cycle} as simulated in Simulink (ode15s solver) using the derived model equations \eqref{eq:FSH}-\eqref{Psynt} and parameters listed in Tab. \ref{table1}.}
  \label{fig:hormones}
\end{figure}

The FSH, LH, E2 and P4 graphs shown in Fig. ~\ref{fig:hormones} result from simulation of the derived equations (1)-(8) using Simulink in Matlab (R2025b) with the ode15s solver. Parameters have been estimated in order to not only obtain similar hormone dynamics but also match the concentration and activity values reported in \cite{welt1999female} (Fig. \ref{fig:scheme}). The parameters used for simulation are listed in Tab. \ref{table1}.

\begin{table}[H]
    \centering
    \caption{Parameters of the hypothetical patient \\}
    \begin{tabular}{p{7mm} c || p{7mm} c}
        \hline\hline
        \textbf{Param.} & \textbf{Chosen Value} & \textbf{Param.} & \textbf{Chosen Value}\\
        \hline
        $b_{\mathrm{F}}$ &  1 IU $\mathrm{L}^{-1}$ $\mathrm{days}^{-1}$ & $b_{\mathrm{L}}$ & 0.001 IU $\mathrm{L}^{-1}$ $\mathrm{days}^{-1}$ \\
        $\alpha_{\mathrm{F}_1}$ & 12 IU $\mathrm{L}^{-1}$ $\mathrm{days}^{-1}$ & $\alpha_{\mathrm{L}_1}$ & 7 IU $\mathrm{L}^{-1}$ $\mathrm{days}^{-1}$\\
        $\alpha_{\mathrm{F}_2}$ & 45 IU $\mathrm{L}^{-1}$ $\mathrm{days}^{-1}$ & $\alpha_{\mathrm{L}_2}$ & 600 IU $\mathrm{L}^{-1}$ $\mathrm{days}^{-1}$\\
        $\alpha_{\mathrm{F}_3}$ & 0.8 IU $\mathrm{L}^{-1}$ $\mathrm{days}^{-1}$ & $\alpha_{\mathrm{L}_3}$ & 7 IU $\mathrm{L}^{-1}$ $\mathrm{days}^{-1}$\\
        $K_{\mathrm{F}_1}$ & 20 $\mu$g $\mathrm{L}^{-1}$ & $K_{\mathrm{L}_1}$ & 70 $\mu$g $\mathrm{L}^{-1}$\\
        $K_{\mathrm{F}_2}$ & 300 $\mu$g $\mathrm{L}^{-1}$ & $K_{\mathrm{L}_2}$ & 300 $\mu$g $\mathrm{L}^{-1}$\\
        $K_{\mathrm{F}_3}$ & 2 $\mu$g $\mathrm{L}^{-1}$ & $K_{\mathrm{L}_3}$ & 70 $\mu$g $\mathrm{L}^{-1}$\\
        $i$ & 2 & $j$ & 4\\
        $k_{\mathrm{F}}$ & 0.78 IU $\mathrm{L}^{-1}$ $\mathrm{days}^{-1}$  & $k_{\mathrm{L}}$ & 2.1 IU $\mathrm{L}^{-1}$ $\mathrm{days}^{-1}$  \\ \hline
        
        $\lambda_1$ & 2.4 $\mathrm{days}^{-1}$ & $d_1$ &$0.05$ $\mathrm{days}^{-1}$\\
        $\lambda_2$ & 2.4 $\mathrm{days}^{-1}$ & $d_2$ &$0.05$ $\mathrm{days}^{-1}$\\
        $\lambda_3$ & 2.4 $\mathrm{days}^{-1}$ & $d_3$ &$0.05$ $\mathrm{days}^{-1}$\\
        $\lambda_4$ & 2.4 $\mathrm{days}^{-1}$ & $d_4$ &$0.05$ $\mathrm{days}^{-1}$\\\hline
        
        $r_{\mathrm{A}}$& 5 $\mathrm{days}^{-1}$ & $L_{\mathrm{lim}}$ & 142 IU $\mathrm{L}^{-1}$ \\
        $d_{\mathrm{R}}$& 0.05 $\mathrm{days}^{-1}$ \\
        $r_{\mathrm{L}}$ & 0.045 $\mathrm{days}^{-1}$& $d_{\mathrm{L}}$ & 0.05 $\mathrm{days}^{-1}$\\ \hline

        $b_{\mathrm{E}}$ & 20 $\mu$g $\mathrm{L}^{-1} \mathrm{days}^{-1}$ & $k_{\mathrm{E}}$ & 1.2 $\mu$g $\mathrm{L}^{-1} \mathrm{days}^{-1}$ \\
        $\alpha_{\mathrm{E}_1}$ & 420 $\mu$g $\mathrm{L}^{-1} \mathrm{days}^{-1}$ & $\alpha_{\mathrm{E}_2}$ & 300 $\mu$g $\mathrm{L}^{-1} \mathrm{days}^{-1}$ \\
        $K_{\mathrm{E}_1}$ & $2 \cdot 10^7$ cells & $K_{\mathrm{E}_2}$ & $1.7*10^7$ cells \\ \hline

        $r_{\mathrm{V}}$ & 0.35 $\mathrm{days}^{-1}$ & $d_V$ & 0.004 $\mathrm{days}^{-1}$ \\
        $\gamma_{\mathrm{P}}$ & 1.6038 ng $\mathrm{L}^{-1}$\\ \hline 
        \hline \hline
    \end{tabular}
    
    \vspace{-0.5cm}
    \label{table1}
\end{table}

\vspace{5mm}

First and foremost, trends from published data in \cite{welt1999female}, which are depicted in Fig. \ref{fig:scheme}, are reproduced by the proposed model. Furthermore, the obtained graphs shown in Fig. \ref{fig:hormones} indicate that the values at the beginning equal the values at the end of a cycle for all four hormones. This is an important outcome of the simulation, as a new cycle is supposed to start after the end of a cycle and the model seems to enable the periodicity.

\vspace{5mm}

In particular for the first half of the cycle, FSH trends from \cite{welt1999female} (depicted in Fig. \ref{fig:scheme}) could be well reproduced in the model simulation. The first peak in early follicular phase as well as the mid-cycle peak slightly before ovulation can be observed in the graph (first panel of Fig. \ref{fig:hormones}). In luteal phase, a minor third peak can be observed which has not been reported in \cite{welt1999female}, but is visibile in patient-specific fluctuations observed in other real-life data \cite{francis2024quantitative}. Importantly, the FSH levels increase towards initial values at the end of the 28-days period, which is important for follicle recruitment and, hence, the start of the following cycle \cite{bhartiya2021overview}. 

As to LH trends, the mid-cycle peak and levels close to zero during all other cycle phases could be reproduced by the model. The depictions of the data published in \cite{welt1999female} (Fig. \ref{fig:scheme}) link the LH peak to ovulation. However, a slight temporal shift is commonly assumed and it has been reported that ovulation occurs approximately 9 hours after the LH peak \cite{pauerstein1978temporal}. The LH peak occurring slightly prior to ovulation in our model is representing this small difference in timing of LH peak and ovulation. Importantly, LH levels remain greater than zero after the peak, which is indispensable for previously discussed effects of LH in the luteal phase, e.g., concerning luteinization (Sec. \ref{sec:cycle}).

The E2 dynamics shown in Fig. \ref{fig:scheme} \cite{welt1999female} are represented quite accurately by the model: The preovulatory peak at a concentration of about 300 pg/mL as well as a second peak in the luteal phase can be identified in the graph (Fig. \ref{fig:hormones}). The data from \cite{welt1999female} suggests two peaks during luteal phase (Fig. \ref{fig:scheme}), but there seems to be individual variability and possibly a dependence on data generation and evaluation, as data from \cite{verdonk2019estradiol} show only one post-ovulatory E2 peak when averaging data from 30 individuals.

The P4 graph shows trends similar to those shown in Fig. \ref{fig:scheme}, i.e., P4 concentrations start increasing mid-cycle and peak at approximately day 19-20, before decreasing again. The start of P4 synthesis is already slightly before ovulation, more precisely, at the LH peak which is biologically sound \cite{devoto2009human}. The decrease yielded by the simulation is steeper than reported in the data which facilitates a final level of approximately zero at the end of the cycle. Since only one cycle was considered in \cite{welt1999female}, variablility within the data might yield that P4 levels at the end of the cycle do not exactly equal the levels at the beginning of the cycle. The proposed model, however, is intended to be capable of representing the periodicity of the process at the cost of a steeper decline.

\begin{table}[H]
\caption{Initial conditions of integrators}
    \centering
    \begin{tabular}{c c || c c}
        \hline\hline
        \textbf{Integrator} & \textbf{Initial condition} & \textbf{Integrator} & \textbf{Initial condition}\\
        \hline
        FSH & 10 mIU/mL & LH & 3 mIU/mL \\
        E & 10 ng/mL \\ \hline

        $G_1$ & $5 \cdot 10^5$ cells & $G_2$ & 0 cells \\
        $S$ & 0 cells & $M$ & 0 cells \\ \hline

        $R$ & 0 cells & $L$ & 0 cells\\
        $V$ & 0.1 \% \\
        
        \hline \hline
    \end{tabular}
    
    \vspace{-0.5cm}
    \label{tab:initials}
\end{table}

\vspace{5mm}

In order to keep the model structurally simple, another limitation had to be allowed. A temporal shift of the mid-cycle E2 and FSH/LH peaks is commonly assumed and has been reported in \cite{pauerstein1978temporal} as well. The model does not reproduce this shift, since steeply increasing FSH and LH levels are tightly coupled to high E2 levels due to the according feedback mechanism (described in detail in Sec. \ref{subsec:gonadotropins}). A more accurate timing and representation of the biological process could be obtained, e.g., by linking the FSH and LH peaks to the maximal slope of the E2 curve. Since this would increase the mathematical complexity of the model, this was however relinquished. Moreover, the peaks are approximately located at the same time in \cite{welt1999female}, which indicates a certain individual variablility in timing as well.

\section{Conclusion and outlook}

In this work, a mechanistic model of the human menstrual cycle has been developed. More precisely, the monthly courses of the four cycle-driving hormones FSH, LH, E2 and P4 were represented in a 10-dimensional ODE-model with 40 parameters. Despite the mathematical and structural simplicity, simulation indicates that the model is suitable to reproduce trends from published hormone activity and concentration data. Thanks to the absence of delay differential equations and a comparably low number of parameters, the simulation of the proposed model is not computationally costly. Therefore, the model is a concise representation of the hormonal courses during the menstrual cycle which can be used as a low-cost systemic background for models of other processes such as embryo implantation and endometrial recovery. Future work on this model aims at further exploring these processes. Furthermore, parameters of the model have been estimated to reproduce average trends from published data. In future research, these parameters could be fit to data of individual patients in order to obtain a representation of their hormone levels. This personalized model could then be used for further simulations, e.g., in pharmacological experiments (such as treatment decisions or dose optimization) or in \textit{in-vitro}-fertilization treatment planning.

\bibliographystyle{IEEEtran}
\bibliography{references}




\end{document}